\documentstyle[12pt,epsfig]{article}
\begin{document}
${}$
\vspace{3cm}
\begin{center}
{LONG--DISTANCE CONTRIBUTION TO EXCLUSIVE RARE DECAYS\\
OF HEAVY MESONS IN QUARK MODEL.\\
${}$\\
${}$\\
Dmitri Melikhov\\
\it Nuclear Physics Institute, Moscow State University, Moscow, 119899, Russia}
\end{center}

\vspace{5cm}
\baselineskip=12pt

{\small 
Nonperturbative effects in exclusive reactions $B\to K^{(*)}ll$ are discussed. 
Form factors which describe the main long--distance contribution to the meson
transition amplitude are calculated within the dispersion 
formulation of the quark model: namely, the form factors in the decay region 
are 
expressed as relativistic double spectral representations through the wave 
functions of the initial and final mesons. 
The dilepton forward--backward 
asymmetry and the lepton polarization 
turn out to be largely independent of the particular choice of the quark 
model parameters and can be predicted with high accuracy. 
At the same time, these asymmetries are sensitive to the short--distance
structure of the theory and might be used as a test of the Standard Model and 
probe of new physics.}

\vspace{3cm}

\begin{center}
{\it to appear in the Proceedings of the XXXIInd Rencontres de Moriond\\
'Electroweak Interactions', Les Arcs, France, March 1997.}
\end{center}

\baselineskip=15pt

\newpage
Rare decays of $B$ mesons induced by the flavour--changing neutral current
(FCNC)
transition $b\to s,d$ provide an important probe of the Standard Model (SM) 
and its extentions. These decays are forbidden at the tree level  
and occur in the lowest order through one--loop diagrams and thus open a
possibility to probe at comparatively low energies the structure of the theory 
at high mass scales which shows through virtual particles in the loops. 
Thus deviations from the SM might be observed in FCNC decays long before a
direct observation of the new particles. On the other hand, measurements of the
FCNC processes provide bounds on the numerical values of the parameters of the
models beyond the Standard Model (SM). 

In order to reliably separate the short--distance effects which contain the information 
on the short-distance structure of the theory, nonperturbative 
long--distance contributions which enter the amplitudes of the exclusive rare $B$
decays should be known with sufficient accuracy. 
Theoretical study of these contributions encounters the problem of describing the
hadron structure and this gives the main uncertainty in predictions for exclusive 
rare decays. 

A strategy of the analysis of exclusive semileptonic $B$ decays looks as 
following:
It is convenient to describe the $b\to s$ transition within the framework of the
low--energy effective theory. Integrating out the heavy degrees of freedom one 
arrives at the effective Hamiltonian at the matching scale $\mu\simeq M_W$ 
which has the form ${}^{1)}$
\begin{equation}
\label{heff}
{\cal H}_{eff} = -\frac{4G_F}{\sqrt{2}} V_{tb} V_{ts}^\ast\, \sum_i
C_i(\mu) \, O_i(\mu). 
\end{equation}
$O_i$ are the basis operators and it is the set of the the Wilson coefficients 
that $C_i(\mu)$ encodes all information on short--distance physics. 

Next, it is necessary to go down to the hadronic scale relevant to the meson transition 
$\mu\simeq m_b$ in order to avoid large logarithms in the meson transition amplitudes. 
The QCD evolution of the Wilson coefficients from $\mu=M_W$ to $\mu=m_b$ is given by the 
matrix of the anomalous dimensions of the the basis operators $O_i$ ${}^{2)}$.
The 4--quark operators from the basis $O_i$ generate the long--distance contribution to
the quark transition $b\to sl^+l^-$ which is mainly due to the $J/\psi$ 
and $\psi'$ resonances in the dilepton channel. The effective Hamiltonian for the
transition $b\to sl^+l^-$ with a built--in long--distance contribution reads ${}^{3)}$  
\begin{eqnarray}
\label{heffbtosll}
{\cal H}_{eff}(b\to sl^{+}l^{-})&=&{\frac{G_{F}}{\sqrt2}}{\frac{\alpha_{em}}{2\pi}}
           V^{*}_{ts}\,V_{tb}\left[\,-2i{\frac{m_b}{q^2}}C_{7}(m_b)
           ({\bar s}\sigma_{\mu\nu}q^{\nu}(1+\gamma_{5})b)({\bar
           l}\gamma^{\mu}l)\right.\\
           &+&\left. C_{9}^{eff}(m_b,q^2)({\bar s}\gamma_{\mu}
           (1-\gamma_{5})b)({\bar l}\gamma^{\mu}l)+
           C_{10}(m_b)({\bar s}\gamma_{\mu}(1-\gamma_{5})b)({\bar
           l}\gamma^{\mu}\gamma_{5}l)\right]  \nonumber
\end{eqnarray}
where $C_{9}^{eff}(m_b,q^2)$ involves the effects of $\psi$ and $\psi'$. 

The amplitude of the reaction $B\to K^{(*)}l^{+}l^{-}$ is given by the meson matrix 
element of the operator eq.(\ref{heffbtosll}). Long--distance dynamical effects 
related to meson formation are contained in relativistic--invariant form factors 
which appear in Lorentz covariant expansion of the meson amplitudes of bilinear quark
currents $\bar b \Gamma s$ from the effective Hamiltonian (\ref{heffbtosll}). 

Various nonperturbative theoretical approaches have been used for 
calculating the form factors of rare semileptonic $B\to K^{(*)}$ decays: 
light--cone quark model ${}^{4)}$, 
constituent quark picture ${}^{5)}$, 
heavy--quark symmetry (HQS) relations ${}^{6,7)}$, 
QCD Sum Rules (SR) ${}^{8,9)}$. 

The results of these nonperturbative considerations yield quite uncertain predictions 
for branching ratios of these decays  
in the SM ${}^{3,4,7-10)}$ 
which hamper isolation 
of short--distance effects from exclusive rare $B$ decays. 

We have studied the transition form factors within a dispersion formulation 
of the quark model (QM) ${}^{11)}$. This formulation is based on 
representing the form factors 
as double spectral representations in the channels of the initial and final 
$q\bar q$ pairs through the wave functions of the initial and final mesons. 
We start with the region $q^2<0$ where the double spectral densities of the 
form factors can be calculated from the Feynman graphs. The form factors at
$q^2>0$ relevant to the decay processes are obtained by performing the 
analytical continuation in $q^2$. This procedure yields the appearance of an
anomalous contribution which rises with $q^2$ and completely determines the form
factor at the zero recoil point. Thus the dispersion formulation of the QM
allows a direct calculation of the form factors at $q^2>0$ through the wave
functions of the initial and final mesons once a proper 
spectral representation at spacelike $q$ is constructed. 

However, it should be taken into account that the dispersion 
approach calculates only double spectral density but does not determine 
possible subtraction terms. To specify such terms we refer to the ideas of the
HQ expansion: namely, we require the structure of the $1/m_Q$ expansion 
of the QM form factors in the case of meson transition induced by the heavy quark
transition to be consistent with the structure of meson transition amplitudes
obtained within the Heavy Quark Effective Theory (HQET). 
This comparison shows that no subtractions are
necessary for the transition between pseudoscalar mesons, whereas 
some of the form factors related to the pseudoscalar--to--vector meson
transition require subtractions. This is mainly explained by problems with
constructing a purely $S$--wave vector state in relativistic theory. 

As a result, we arrive at the form factors with the following properties ${}^{12)}$: 
for the transition induced by the 
heavy--to--heavy quark transition they satisfy not only 
the leading--order Isgur--Wise relations ${}^{13)}$ but also 
subleading $O(1/m_Q)$ relations of the HQET ${}^{14)}$ provided the wave 
functions of heavy mesons are localized near the $q\bar q$ threshold with the 
width of order $\Lambda_{QCD}$.  
For the meson decay induced by the heavy--to--light 
quark transition the QM form factors satisfy the leading--order relations between the form 
factors of the vector and tensor currents ${}^{15)}$ .
 
The numerical analysis of the transition form factors for various sets of the 
QM parameters performed in ${}^{16)}$ exhibits a strong dependence of the form 
factors and decay rates on a choice of such parameters. This yields considerable errors in the QM 
predictions for the decay rates (see Table 1). 

{\small
\begin{table}[htb]
\noindent \baselineskip=12pt
{\bf Table 1.} Non-resonant branching ratios of 
rare radiative and semileptonic $B$-decays. Theoretical predictions
are given in units $|V_{ts} / 0.033|^2$. The results of Ref.${}^{8)}$
have been recalculated replacing the value 
$|V_{ts} / 0.04|^2$ with $|V_{ts} / 0.033|^2$. The uncertainties in $V_{ts}$
are not included in the error bars.

\begin{center}
\begin{tabular}{||l||c|c|c||c||} \hline
Ref.                    & QM${}^{16)}$ & HQS${}^{3)}$ & SR${}^{8)}$ & Exp. \\ 
\hline                  
Decay                   & ${\cal{BR}}$        & ${\cal{BR}}$
                        & ${\cal{BR}}$        & ${\cal{BR}}$\\
                        \hline \hline
$B \to K^* \gamma$      & $(3.9 \pm 1.7) \times 10^{-5}$
                        & $(4.9 \pm 2.0) \times 10^{-5}$ & $-$
                        & $(4.2 \pm 1.0) \times 10^{-5}$
                        [17] \\ \hline \hline
$B \to K \ell^+ \ell^-$ & $(4.2 \pm 0.9) \times 10^{-7}$
                        & $(4.0 \pm 1.5) \times 10^{-7}$
                        & $2 \times 10^{-7}$
                        & $< 0.9 \times 10^{-5}$ [18]\\ \hline\hline
$B \to K^* e^+ e^-$                & $(1.4 \pm 0.5) \times 10^{-6}$     
                                   & $(2.3 \pm 0.9) \times 10^{-6}$
                                   & $0.7 \times 10^{-6}$
                                   & $ <1.6 \times 10^{-5}$ [18]
                                   \\ \hline
$B \to K^* \mu^+ \mu^-$            & $(1.0 \pm 0.4) \times 10^{-6}$     
                                   & $(1.5 \pm 0.6) \times 10^{-6}$
                                   & $0.7 \times 10^{-6}$
                                   & $< 2.5 \times 10^{-5}$ [19]
                                   \\ \hline
\end{tabular}
\end{center}
\end{table}
}

\baselineskip=15pt
Fortunately, the dependence of the
asymmetries on the QM parameters is negligible and the forward--backward
asymmetry as well as lepton polarization asymmetry are predicted with high accuracy
in largely model--independent way. 

At the same time the asymmetries in the nonresonance regions 
are sensitive to the values of the Wilson coefficients and thus can be used 
as a test of the SM and a probe of new physics. Fig.1 shows asymmetries 
obtained within the SM and MSSM for 
various regions of the MSSM parameter space ${}^{20)}$.  

\begin{figure}[htb]
\begin{center}
\begin{tabular}{cc}
\mbox{\epsfig{file=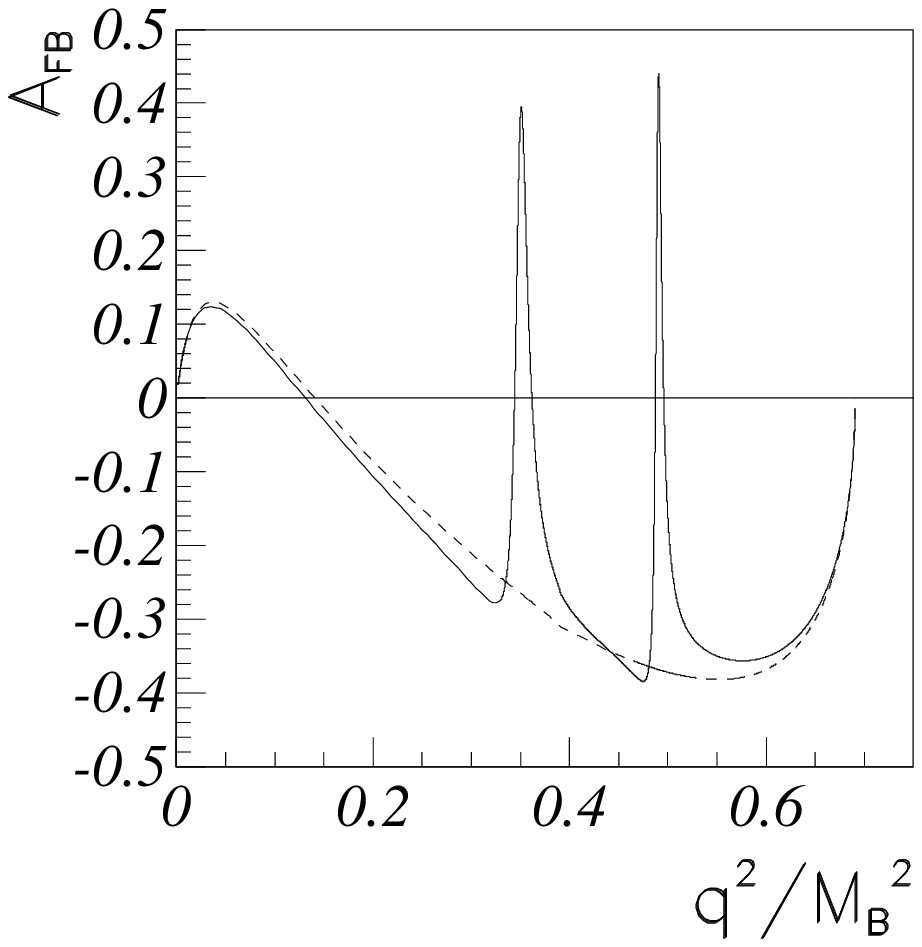,width=7cm}}& 
\mbox{\epsfig{file=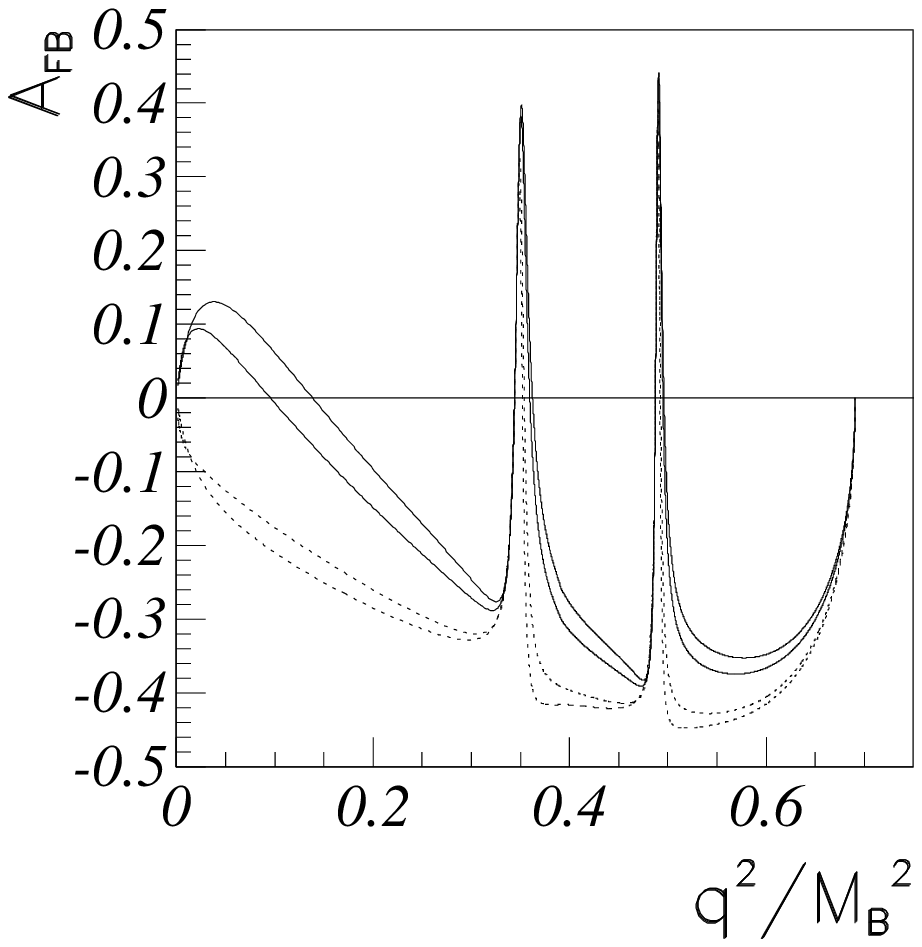,width=7cm}}\\
\end{tabular}\end{center}
\caption{ 
\baselineskip=10pt
Forward--backward asymmetry in $B\to K^*\mu^+\mu^-$. 
(a) in the SM: total -- solid, nonresonance -- dashed.  
(b) in the MSSM for various values of $R_7=C_7(M_W)_{MSSM}/C_7(M_W)_{SM}$: 
upper solid line -- $R_7=1.2$, lower solid line -- $R_7=0.4$; 
upper dashed line -- $R_7=-2.4$, lower dashed line -- $R_7=-4.2$. The regions
between the solid and dashed lines, respectively correspond to the allowed
regions of the MSSM parameter space ${}^{20)}$.  
Because of a weak sensitivity 
of $A_{FB}$ to $C_9$ and $C_{10}$, the latter are equated to their SM values. }
\end{figure}
\baselineskip=15pt

Notice that obtaining more accurate predictions for the
exclusive distributions is also necessary in particular for the extraction of 
$V_{tb}$. To this end one needs a relevant choice of the QM parameters. 
Further analysis of the semileptonic decays and new more accurate data on 
exclusive radiative decay $B\to K^*\gamma$ will be very helpful.  

I take pleasure in thanking J.Tran Thanh Van, J.-M.Frere, and other Organizers for
their warm hospitality, N.Nikitin, S.Simula, B.Stech and K.Ter--Martirosyan for 
useful discussions and RFBR for financial support of this work under grant 
96--02--18121a.
\vspace{.5cm}

\baselineskip=12pt
{\small 
\hspace{2pt} [1] B.Grinstein, M.B.Wise and M.J.Savage, Nucl.Phys. {\bf B319} (1989) 271.

\hspace{2pt} [2] A.Buras and M.M\"unz, Phys.Rev. {\bf D52} (1995) 186.

\hspace{2pt} [3] A. Ali, preprint DESY 96-106. 

\hspace{2pt} [4] W.Jaus and D.Wyler, Phys. Rev. {\bf D41} (1990) 3405.

\hspace{2pt} [5] B.Stech, Phys.Lett. {\bf B354}(1995)447 and hep-ph/9608297.

\hspace{2pt} [6] G.Burdman, Phys.Rev. {\bf D52} (1995) 6400.

\hspace{2pt} [7] W. Roberts, Phys.Rev. {\bf D54} (1996) 863. 

\hspace{2pt} [8] P.Colangelo, F.De Fazio, P.Santorelli and
 E.Scrimieri, Phys.Rev. {\bf D53} (1996) 3672.
 
\hspace{2pt} [9] T.M. Aliev, A.Ozpineci and M.Savci, hep-ph/9612480 and
 hep-ph/9702209.
 
[10] C.Q.Geng and C.P.Kao, Phys.Rev. {\bf D54} (1996) 5636.

[11] D. Melikhov, Phys.Rev. {\bf D53} (1996) 2460; Phys.Lett. {\bf B380} (1996)
363, {\bf B394} (1997) 385.

[12] D.Melikhov, N.Nikitin, hep-ph/9609503. 

[13] N.Isgur and M.B. Wise, Phys.Lett. {\bf B232} (1989) 113;
{\bf B237} (1990) 527.

[14] M. Luke, Phys. Lett. {\bf B252} (1990) 447.

[15] N.Isgur and M.B.Wise, Phys.Rev. {\bf D42} (1990) 2388.

[16] D.Melikhov, N.Nikitin, and S.Simula, hep-ph/9704268. 
 
[17] CLEO collaboration (R.Ammar et al), CLEO CONF 96-05 (1996).

[18] T. Skwarnicki, hep-ph/9512395. 
 
[19] CDF collaboration (T.Speer et al.), FERMILAB CONF-96/320-E (1996). 
 
[20] P.Cho, M.Misiak, and D.Wyler, Phys.Rev.{\bf D54} (1996) 3329.}

\end{document}